\documentstyle[12pt]{article}
\def\reference{\parskip 0pt\par\noindent\hangindent 0.5 truecm}

\begin{document}
%
%
\title{Multi-wavelength behaviour of III~Zw2}
%


\author{Nikita Salvi,  
 Mat J. Page, 
 Jason A. Stevens,  \\  
 Keith O. Mason and 
 Kinwah Wu 
} 

\date{}
\maketitle

{\center
 Mullard Space Science Laboratory, University College London, \\ 
 Holmbury St.~Mary, Dorking, Surrey RH5 6NT, United Kingdom 
 \\njs@mssl.ucl.ac.uk\\[3mm]
}

%
\begin{abstract}

III~Zw2 was observed with XMM-Newton in July 2000. 
  Its X-ray spectrum can be described 
  by a powerlaw of photon index $\Gamma \approx 1.7$ 
  with a Gaussian line at 6.7~keV. 
There is no significant evidence of intrinsic absorption within the source 
  or of a soft X-ray excess.
Multi-wavelength light curves over a period of 25 years 
  show related variations from the radio to the X-rays. 
We interpret the radio to optical emission 
  as synchrotron radiation, self-absorbed in the radio/millimeter region, 
  and the X-rays as mainly due to Compton up-scattering 
  of low-energy photons by the population of high-energy electrons 
  that give rise to the synchrotron radiation.

\end{abstract}


\bigskip 

\section{Introduction}

 
III~Zw2 (PG 0007+106, Mrk 1501) is a Seyfert I galaxy   
  (Arp 1968; Khachikian \& Weedman 1974; Osterbrock 1977) 
  with z = 0.089 (de Robertis 1985). 
Superluminal motion of radio emitting plasma has been observed recently 
  in the source, 
  and this is the first detection of its kind in a spiral galaxy 
  (Brunthaler et al.~2000).  
The source has long been known 
  to show large-amplitude flux variations  
  in the radio (Wright et al.~1977; Schnopper et al.~1978; Landau et al.~1980; 
  Aller et al.~1985; Falcke et al.~1999) and the  
  optical (Lloyd 1984; Clements et al.~1995). 
Variations of smaller amplitude (of less than 50~$\%$), have also been detected
   in the IR (Lebofsky \& Rieke 1980; Sembay, Hanson \& Coe 1987)
  and the UV (Chapman, Geller \& Huchra 1985).   
The X-ray temporal behaviour is less well studied, 
  but comparison of observations at different epochs  
  hints that the X-ray flux might vary substantially.   
The X-ray spectra of III~Zw2, 
  obtained by SAS-3 (Schnopper et al.~1978), Ariel VI (Hall et al.~1981) 
  and Einstein SSS (Petre et al.~1984),  
  can be described by a powerlaw of photon indices $\Gamma \sim 1.3 - 1.7$, 
  with neutral absorption consistent with that of the Galactic column.      
  
\section{X-ray spectrum}

\begin{figure}[ht!]
\begin{center}
\vspace*{6cm}
\leavevmode
\includegraphics{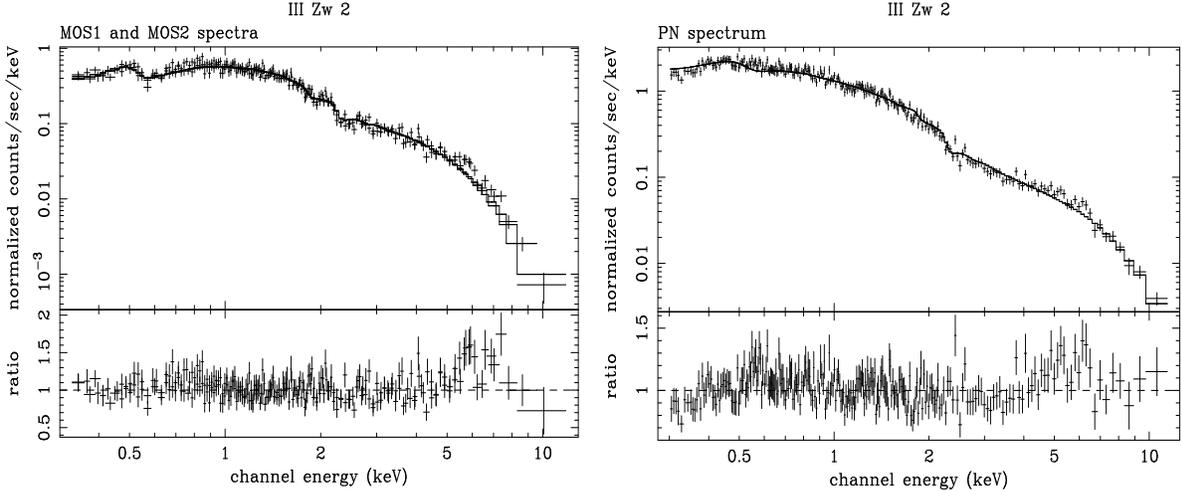} 
\caption{
  Left panel shows a simultaneous powerlaw fit to the MOS1 and MOS2 spectrum 
  ($\Gamma= 1.73\pm0.01$, $\chi^2$/dof = 295/253). 
  Right panel shows a powerlaw fit to the PN spectrum 
  ($\Gamma= 1.75\pm0.01$, $\chi^2$/dof = 317/226). 
  The galactic column was fixed at $N_{\rm H} = 5.72\times10^{20}$ cm$^{-2}$ 
     in both cases. 
  We caution that the PN has a calibration uncertainty of about 150~eV. }
\label{mosspec}
\end{center}
\end{figure}

We observed III~Zw2 with XMM-Newton (Jansen et al.~2001) on 2000 July 7.  
(Details of the observations and data analysis will be presented elsewhere.) 
Figure.~\ref{mosspec} shows the spectra (in the observer frame) 
  from the two EPIC CCD instruments aboard the satellite 
  (MOS spectra in the left panel and PN spectrum in the right panel).  
We fit the spectra initially with a powerlaw and a fixed galactic column of 
  $N_{\rm H} = 5.72\times10^{20}$ cm$^{-2}$. 
The MOS spectra give a photon index of $\Gamma = 1.73\pm0.01$,  
  and a similar value is obtained for the PN spectrum,    
  which is consistent with previous observations. 
Excess emission, however, remains at energies $\sim 5-7$~keV.  
Adding a Gaussian (at E = 6.76 $\pm 0.22$~keV in the rest frame of the source) 
  to the powerlaw improves the fit to the MOS data significantly 
  (with $\Delta \chi^2 $= 45.8 for 3 parameters),  
  suggesting the presence of a broad iron line 
  (or perhaps blended iron lines), with an EW $\sim$ 1.5~keV.  
There is no evidence for a soft X-ray excess. 
The absorption can be explained by the line-of-sight Galactic absorption, 
  consistent with the finding of previous observations 
  (Hall et al.~1981; Petre et al.\ 1984; Kaastra \& de Korte 1988).  
 
\begin{figure}
\begin{center}
\vspace*{20cm}
\leavevmode
\includegraphics{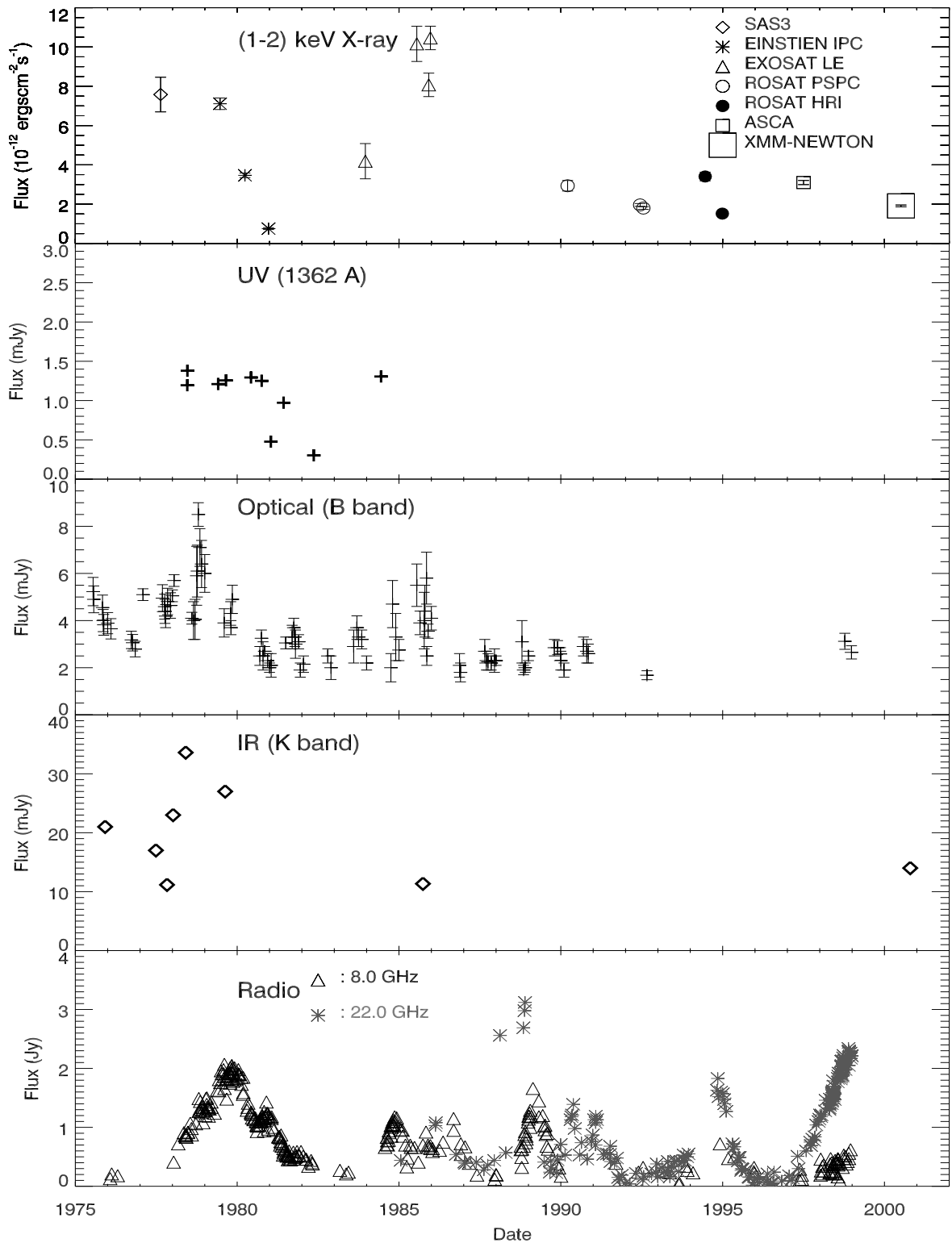} 
\caption{ 
  Multi-wavelength light curves of III~Zw2 showing flux variations, 
     with increasing wavelengths from top to bottom. 
     (See \S3 for the sources of data.) }
\label{multi_lc}
\end{center}
\end{figure}

\section{Multi-wavelength variability} 

Figure.~\ref{multi_lc} shows the multi-wavelength light curves of III~Zw2 
  from 1975 to 2000. 
The X-ray data, with the exception of the XMM observation, 
  are from the HEASARC archives.  
Non-imaging X-ray observations, with the exception of the SAS-3 flux of Aug 1977
  (Schnopper et al.~1978), have been excluded from our analysis due to the
  presence of a bright X-ray source approximately 12 arc minutes from
  III~Zw2. This source may have been responsible for the contamination of the
  EXOSAT ME flux measurements of III~Zw2 in 1985 (see Tagliaferri et al.~1988).  
As the spectral photon indices  
  are similar in the XMM observation and in previous X-ray observations 
  (e.g. Hall et al.~1981; Petre et al.~1984),  
  we have used the powerlaw index $\Gamma = 1.7$ 
  (deduced from the XMM EPIC data, \S2)
  to determine the $1-2$~keV flux for all observations. 
The optical light curve is a combination of CCD data of the La Palma archives  
  and various photographic monitoring observations  
  (Lloyd 1984; Clements et al.~1995). 
The UV data are from the IUE archives (see also Chapman et al.~1985), 
  the IR data are compiled from various publications 
  (Neugebauer et al.~1979; Lebofsky \& Rieke 1980; Condon et al.~1981; 
  Hyland \& Allen 1982; McAlary et al.~1983; Edelson \& Malkan 1987; 
  Sembay, Hanson \& Coe 1987), 
  and the radio data at 22~GHz are from Mets\"{a}hovi 
  (H.~Terasranta, private communication) 
  and at 8~GHz are from Michigan 26m
  (M.~F.~Aller, private communication). 

The radio and optical wavelengths are better monitored (with optical up to
1990) than the other wavelengths. The source shows 10- and 20-fold variations
in X-rays and in radio respectively over time-scales of years. The amplitude of
variations is relatively smaller in optical
and IR. The optical and X-ray observations seem to be correlated, but the
correlation is less certain between other wavelengths. Comparison of flares
at the optical and radio wavelengths indicate the possibility that the 8-Ghz
radio flux peaks later than the optical flux (Clements et al.~1995). However,
the sampling of the data is not good enough to quantify the exact time lead/lag
between flares at other wavelengths.

\section{Spectral energy distribution}

\begin{figure}[ht!]
\begin{center}
\vspace*{6cm}
\leavevmode
\includegraphics{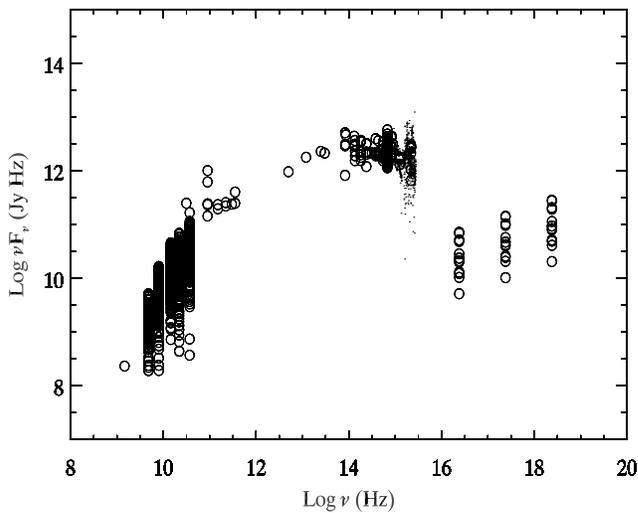} 
\caption{ 
  Spectral energy distribution for III~Zw2. 
  It includes radio to X-ray data taken over the last 25 years. 
  (See \S3 for the sources of data.) }
\label{multispec}
\end{center}
\end{figure}

Figure.~\ref{multispec} shows the spectral energy distribution of III~Zw2,
  incorporating data from 1975 to 2001. 
The lack of spread in data in the frequency range $10^{11}-10^{14}$~Hz 
  is probably due to insufficient sampling in that region. 
The spectrum peaks at the optical/IR wavelengths,   
  with no evidence for an excess in UV and soft X-rays.
The absence of a thermal blackbody component in the broad-band spectrum casts 
  doubts on a model in which a substantial proportion of UV/X-rays are emitted
  from an optically thick accretion disk (cf. Kaastra \& de Korte 1988).
The large-amplitude flare observed in all wavelengths around 1980 suggests the
  possibility that the emission from  X-rays to radio has a
  common origin. 
One possibility is that the broad-band emission from 10$^9$-10$^{14}$ Hz is due
  to synchrotron radiation from a cloud of relativistic electrons.
The synchrotron emission would be self-absorbed in the radio/millimeter region
  and would have a high-energy cutoff at wavelengths shortward of the optical.

The X-rays are, however unlikely to be direct synchrotron emission. 
From the coincidence of X-ray flares with those in the other
wavelengths (observed in the eighties, especially in optical), we interpret the X-rays as
Compton up-scattered emission, by high energy relativistic electrons that give rise to
radio and optical/IR synchrotron radiation.
High time resolution, multi-wavelength polarization observations are required to
verify  whether or not the broad-band radio and optical radiation is of
synchrotron origin.

%





\section*{Acknowledgments}
We thank H.~Terasranta and M.~F.~Aller for the 22-GHz and 8-GHz radio data 
  respectively. NJS acknowledges receipt of a PPARC studentship.

\section*{References}






\reference 
Aller, H. D., Aller, M. F., Latimer, G. E., \and Hodge, P. E. 1985, 
  ApJS, 59, 513 
\reference   
Arp, H. 1968, ApJ, 152, 1101 
\reference  
Brunthaler, A., et al. 2000, A\&A, 357, L45 
\reference 
Chapman, G. N. F., Geller, M. J., \and Huchra, J. P. 1985, ApJ, 297, 151
\reference 
Clements, S. D.,  Smith, A. G., Aller, H. D., \and Aller, M. F. 1995, AJ, 
  110, 529   
\reference 
Condon, J. J., O'Dell, S. L., Puschell, J. J., \and Stein, W. A. 1981, 
  ApJ, 246, 624  
\reference 
de Robertis, M. 1985, ApJ, 289, 67 
\reference 
Edelson, R. A., \and Malkan, M. A. 1987, ApJ, 323, 516
\reference 
Falcke, H., et al. 1999, ApJ, 514, L17
\reference 
Hall, R., Rickett, M. J., Page, C. G., \and Pounds, K. A. 1981, Sp.~Sci.~Rev., 
  30, 47 
\reference 
Hyland, A. R., \and Allen, D. A. 1982, MNRAS, 199, 943 
\reference
Jansen, F., et al. 2001, A\&A, 365, L1 
\reference
Kaastra, J. S., \& de Korte, P. A. J. 1988, A\&A, 198, 16  
\reference 
Khachikian, E. Y., \and Weedman, D. W. 1974, ApJ, 192, 581 
\reference 
Landau, R., Epstein, E. E., \and Rather, J. D. G. 1980, AJ, 85, 363 
\reference 
Lebofsky, M. J., \and Rieke, G. H. 1980, Nature, 284, 410    
\reference 
Lloyd, C. 1984, MNRAS, 209, 697 
\reference
McAlary, C. W., McLaren, R. A., McGonegal, R. J., \and Maza, J. 1983, ApJS, 52, 341
\reference 
Neugebauer, G., Oke, J. B., Becklin, E. E., \and Matthews, K. 1979, ApJ, 230, 79 
\reference 
Osterbrock, D. E. 1977, ApJ, 215, 733 
\reference 
Petre, R., Mushotzsky, R. F., Krolik, J. H., \and Holt, S. S. 1984, ApJ, 280, 499 
\reference 
Schnopper, H. W., et al. 1978, ApJ, 222, L91   
\reference 
Sembay, S., Hanson, C. G., \and Coe, M. J. 1987, MNRAS, 226, 137 
\reference 
Tagliaferri, G., et al. 1988, ApJ, 331, L113 
\reference 
Wright, A. E., Allen, D. A., Krug, P. A., Morton, D. C., \and Smith, M. G. 1977, 
  IAUC, 3145, 2


\end{document}